\documentclass[a4paper]{article}
\usepackage[ansinew]{inputenc}
\usepackage{times}
\usepackage[T1]{fontenc}
\usepackage{graphicx}
\usepackage{geometry}
\usepackage{amssymb}
\usepackage{amsmath}

\usepackage{rotating}

\usepackage{xcolor}
\colorlet{shadecolor}{gray!25}
\usepackage{subfig}
\usepackage{float}

\author{Ludwig A. Hothorn,\\ 
Im Grund 12, D-31867 Lauenau, Germany\\ (retired from Leibniz University)\\
Frank Schaarschmidt, \\
Leibniz University Hannover, Faculty of Natural Sciences,\\ Herrenhauser Str. 2, D-30419 Hannover, Germany}

\title{A modified Armitage test\\ for more than a linear trend on proportions.}
\begin{document}

\maketitle
\begin{abstract}
The Armitage test for linear trend in  proportions can be modified using the multiple marginal model approach for three regression models with arithmetic, ordinal and logarithmic dose scores simultaneously, to be powerful against a wide range of possible dose response relationships. Moreover, it can be used for particular designs in the generalized linear (mixed) model for the three common effect sizes odds ratio, risk ratio and risk difference. The related R package \verb|tukeytrend| allows simple generalizations, e.g.  the analysis 2-by-k table data with a possible plateau shape or analysing overdispersed proportions. The evaluation of further real data examples are available in a vignette to that R package.
\end{abstract}

\section{Introduction}\label{sec1}

Armitage's landmark paper \emph{Tests for linear trends in proportions} \cite{Armitage1955} is widely used in the evaluation of   case-control genetic association studies \cite{Zhou2017}, replicated toxicological bioassays \cite{Novelo2017}, clinical dose finding studies \cite{Lilot2013},  exposure epidemiological table data \cite{Arendt2016}, among others. This CA-test is a scores test for weighted linear regression for the differences of proportions, organized in the simple 2-by-k table format with quantitative dose levels $D_i$. For the proportions $\pi_i$ is the common-formulated ordered alternative hypothesis $H_1: \pi_1<\pi_2<,...,<\pi_k | \pi_1<\pi_k$ (against $H_0: \pi_1=\pi_2=,...,=\pi_k$) is somewhat understated. Rather, it is tested exactly whether the slope parameter $b$ of a linear regression differs from zero $H_1:b>0$.
For $k$ binomial variates $y_i$ with response probabilities $p_i$ in a randomized one-way layout with dose levels $D_i$ using samples size $n_i$ the test statistics is:\\ $T_{CA}=\sum_{i=1}^k y_i(D_i-\overline{D})/\sqrt{\overline{p}(1-\overline{p}[\sum_{i=1}^k n_i(D_i-\overline{D})^2]}$\\
which is asymptotically univariate normal distributed. In principle it can be formulated as a two-sided or one-sided test, but it is rare to restrict to a trend alternative, not knowing in which direction. Therefore the one-sided version is considered here only. A Yates-type continuity correction or an exact unconditional and conditional version \cite{Shan2012} are available for small $n_i$ \cite{Nam1987}. \\

One issue with the CA-test is that it is considered as an order restricted test, sensitive to many monotonous alternatives, but is defined for a linear alternative of the quantitative covariate $D_i$ only. This becomes particularly clear in the genetic case-control association tests, where the CA test \cite{Debrah2017} is defined for the additive mode of inheritance only, whereas the additional testing of dominant and recessive modes is more appropriate according to the genetic paradigm \cite{Hothorn2009}.  Of course, a linear regression test shows not only power for exactly the linear alternative, but also in its neighborhood up to sublinear shapes. However, for supralinear profiles like a plateau shape, which is  the recessive mode in genetics, the power is comparatively low. Considering alternatively dose as a qualitative factor, the robust score-independent test \cite{Winell2018} or multiple contrast tests \cite{Bretz2002} can be recommended. \\
 
If the dose is still considered as a quantitative covariate, then the Tukey trend test \cite{Tukey1985} can be used, namely simultaneous considering three regression models for the metameters $D_i$, $rank(D_i)$ and $log(D_i)$, \cite{Schaarschmidt2018}, providing the joint distribution for this maximum test by means of multiple marginal models (mmm) \cite{Pipper2012}. 	This requires the formulation of logistic regression models, but allows extensions in generalized linear (mixed) models - a second issue discussed here. Exactly where modifications of the original CA-test are difficult, e.g. the adjustments against covariates, two-way to hierarchical designs, overdispersed proportions instead of simple table data, multi-center studies, repeated measurements, multiple endpoints, among others. \\

Although tests are divided into those for quantitative and qualitative modeling of dose, our approach allows the simultaneous consideration of regression models ($D_i$ as covariate) \textbf{and} multiple contrasts ($D_i$ as qualitative factor levels) simultaneously, a possible alternative for the controversial use of post-hoc categorized exposure levels \cite{Barnwell-Menard2015}.\\

Not only the common used p-value for the CA-test is provided but also a compatible confidence interval for a well-chosen effect size are available in the line of the recent p-value controversy \cite{Vila2017, Amrhein2019}.\\

Notice, the CA-test is also used to claim temporal increasing (decreasing) incidences, e.g. the decreasing proportion of abnormal cervical smears in Korea cancer screening program from 2009 to 2014 \cite{Shim2017}, although dependent times contradicts the assumption of independent doses. \\

\section{An asymptotic Tukey-type trend test for proportions}
The primary idea of Tukey \cite{Tukey1985} is the simultaneous fitting of $k=(1,2,3)$  low parametrized regression models:\\
(1) $logit(p_i)^{arithmetic}= \alpha_1 + \beta_1 D_i + \epsilon_i$, \\
(2) $logit(p_i)^{ordinal}= \alpha_1 + \beta_2 rank(D_i) + \epsilon_i$\\
(3) $logit(p_i)^{logarithmic}= \alpha_1 + \beta_3 log(D_i) + \epsilon_i$\\
and the selection of the best model by a maximum test. The joint inference for the three slope parameters $\beta_1, \beta_2, \beta_3$ for the three models $m=1,2,3$ for the union-intersection test hypothesis
$H_{0}: \bigcap_{m=1}^{3} \beta_{m} \leq 0 ~\mbox{vs.} ~ H_{A}: \bigcup_{m=1}^{3} \beta_{m} > 0$ can be achieved by the mmm approach \cite{Schaarschmidt2018}.

Marginally fitting the model $k$, yields vectors with the parameter estimates $\hat{\boldsymbol{\delta}}_{k}$,the corresponding standard error estimates $\hat{\boldsymbol{\sigma}}_{k}$, and $(I \times P_{k})$ matrices containing the standardized score contributions evaluated at the ML estimates of the model parameters \cite{Pipper2012, Jensen2015}, denoted by $\boldsymbol{\hat{\Psi}}_{k}$. 
Note that the $M$ estimators $\hat{\theta}_{m}$ may be correlated, and thus test statistics in $\boldsymbol{t} = \left(t_{1},...,t_{M}\right)$ may be correlated: models that use different transformations or parametrization of the dose effect but are applied to the same response variable contribute correlated test statistics, because an overall positive trend in the data will cause several or all of the corresponding test statistics to deviate from 0. Further, models applied to different response variables may yield correlated test statistics if the response variables are correlated. In the following, results of \cite{Pipper2012} will be used to account for the correlation among test statistics: For larges samples, the covariance among the estimators $\boldsymbol{\hat{\theta}}$ obtained from multiple marginal models can be estimated from the row vectors $\boldsymbol{\hat{\psi}}_{n}$ of $\boldsymbol{\hat{\Psi}}$, $\boldsymbol{\hat{V}} = 1/N\sum_{n=1}^{N}\boldsymbol{\hat{\psi}}^{T}_{n}\boldsymbol{\hat{\psi}}_{n}$. Standardizing $\boldsymbol{\hat{V}}$ by its diagonal elements yields an asymptotic estimator of the correlation between the $M$ test statistics of interest, $\boldsymbol{\hat{R}}$. Notice, this maximum-test is asymptotic only.\\

 Using the object-oriented R language the realization of this approach is numerically simple: first fitting a logistic regression by \verb|glm| into an object \verb|m1|, using the library \verb|tukeytrend| to find the joint distribution with the correlations between the three model yielding simultaneous confidence limits (or adjusted p-value) for the slope (with have to be back-transformed into OR using \verb|expit|.)

\section{Modifications}
\subsection{Canonical and non-canonical link function in the GLM}
Approaches for the three common effect sizes for proportions can be formulated: odds ratio (OR), risk ratio (RR) and risk difference 
(RD). The logit is the canonical link in generalized linear model (GLM) yielding to the log-odds ratio of slope as effect size.
For small $n_i$ and/or parameters near to the boundary, e.g. $p_0=0$, the parameter estimates may be biased and the confidence interval do not control the level \cite{Bergtold2018}. Because of the pooling properties of multiple contrast tests \cite{Schaarschmidt2008}, even the model will not converged in such cases. Here an adjustment is recommended, such as \emph{add2} correction \cite{Agresti2011} yielding pseudo-score intervals \cite{Lawson2004}. Because of a one-sided test on slope is of interested, we use here an \emph{add1} approach \cite{Schaarschmidt2013}.\\

Risk difference (RD) estimates are available using the identity link in R's GLM basic function whereas the number of iterations should be large chosen. Risk ratio estimates are available by the package \verb|logbin| using an EM-type algorithm \cite{Marschner2015}.\\

Usually the effect size should be clear a-priori based on design, randomization and style of interpretation. This is not always the case in the p-value-focused world, and therefore a maximum test over these 3 link functions is proposed a trifle provocative. Using the multiple marginal model concept, such a double maximum test is available for transformations $\xi(p_i)$ (and the 3 dose metameters) because the data matrix is the same. As a result you get which model related link function and which dose metameter is most in the alternative. Notice, these models are highly correlated, so the additional conservatism is  comparatively low with respect to the possibility of choosing the best model under error control. \\

This approach is demonstrated by simple 2-by-k-table data for incidence of squamous cell papilloma in male B6C3F1 mice administered 0, 0.0875, 0.175, 0.35, or 0.70 mM acrylamide \cite{Beland2013}:

\begin{table}[ht]
\footnotesize
\centering
\caption{Example 2-by-k-table data} 
\begin{tabular}{rrrrrr}
  \hline
Dose & 0 & 0.0875 & 0.175 & 0.35 & 0.70 \\ \hline
  No. papilloma & 0 & 2 & 2 & 6 & 6 \\ 
  No. mice & 46 & 45 & 46 & 47 & 44 \\ 
   \hline
\end{tabular}
\end{table}

The R-code is relatively simple using the R-package \emph{tukeytrend} here for the log(OR) (see more details in the vignette \cite{vignette}):
\footnotesize
\begin{verbatim}
library(tukeytrend)
swaOR <-glm(cbind(events+0.5,(n-events)+0.5)~dose, data=squam, family= binomial(link="logit"))
tuOR <- tukeytrendfit(swaOR, dose="dose", scaling=c("ari", "ord", "arilog"))
pOR <- summary(glht(model=tuOR$mmm, linfct=tuOR$mlf, alternative="greater"))
\end{verbatim}
\normalsize

For the canonical link the model with min(p) is for the ordinal dose metameter (see  Table \ref{tab:tp6}) and the corresponding OR (and its lower confidence limit) can be easily estimated: $1.70; [1.18,]$. This means that with a dose increase of rank 1, the OR for the tumor incidence increases by at least $118\%$. The optimal model without a-priori model choice (Opt: RD) is  for RD and $log(D_i)$, where the p-value of $0.0013$ is only slightly higher than for the a-priori selected model ($0.00068$).

\begin{table}[ht]
\centering
\footnotesize
\caption{p-values for 4 different models: effect sizes OR, RD and RR, and the best of (OR,RR,RD) } 
\label{tab:tp6}
\begin{tabular}{lll}
  \hline
Effect size & Metameter & Adj. $p$-value \\ 
  \hline
OR & ari & 0.00833 \\ 
  OR & ord & 0.00614 \\ 
  OR & log & 0.00632 \\ \hline
  RD & ari & 0.00280 \\ 
  RD & ord & 0.00069 \\ 
  RD & log & 0.00068 \\ \hline
  RR & ari & 0.00591 \\ 
  RR & ord & 0.00454 \\ 
  RR & log & 0.00441 \\ \hline \hline
  Opt:RD & Opt:log & 0.00131 \\ 
   \hline
\end{tabular}
\end{table}

\subsection{Joint testing dose as quantitative covariate and qualitative factor}
The mmm-approach allows maximum tests over any linear models, e.g. a joint test of Tukey-type and Williams-type trend test \cite{Williams1971}, i.e. a regression-type and an ANOVA-type linear contrast model \cite{Schaarschmidt2018}. Two contradictory aspects meet: increasing the conservativeness of the test by including $3+(k-1)$ single tests (instead of only 3) and an almost balanced power behavior for any of sub-, supra- and linear profiles. The power loss of this joint test against the Tukey-test is tolerable because these many models are highly correlated \cite{Schaarschmidt2018}. It allows the interpretation of elementary confidence intervals (or p-values) for both the slopes and the pooled contrasts. The latter is the essential advantage over a max-test including CA-trend test and Williams-isotonic test \cite{WILLIAMS1977} \cite{Peddada2007}. This joint test is particularly interesting for supra-linear curves, e.g. a plateau effect, as it shows an average high quality for any shape \cite{Schaarschmidt2018}.  As an example the number of cured patients treated with $0, 1, 2, 4 \%$ 
flutrimazol in a clinical dose-finding study in acute vulvovaginal candidosis \cite{Palacio2000} is used:

\begin{table}[ht]
\centering
\scriptsize
\caption{Example 2-by-k-table data with a possible plateau shape} 
\begin{tabular}{rrrrr}
  \hline
 & 1 & 2 & 3 & 4 \\ 
  \hline
Dose & 0 & 1 & 2 & 4 \\ 
  No. cured & 12 & 22 & 22 & 20 \\ 
  No. patients & 33 & 28 & 32 & 31 \\ 
   \hline
\end{tabular}
\end{table}
\normalsize
The GLM-object \verb|lmf| is plugged into the \verb|tukeytrendfit| function, containing both the three regression models and the Williams contrast test \cite{Williams1971}: 

\footnotesize
\begin{verbatim}
lmf <-glm(cbind(events,n-events)~dose, data=flutri, family= binomial(link="logit"))
TW <- tukeytrendfit(lmF, dose="dose", 
scaling=c("ari", "ord", "arilog", "treat"), ctype="Williams")
ETW <- summary(glht(model=TW$mmm, linfct=TW$mlf, alternative="greater"))
\end{verbatim}
\normalsize

While none of the regression models is in the alternative, the William multiple contrast test shows a significant increase in the cure rate for the pooled doses in Table \ref{tab:extw}.

\begin{table}[ht]
\centering
\scriptsize
\caption{Tukey Trend or Williams test (canonical link)} 
\label{tab:extw}
\begingroup\small
\begin{tabular}{l|l|rr}
  \hline
Dose & Model& Test statistics & $p$-value \\ 
  \hline
Quantitatively& Tukey: arithm & 1.7 & 0.14 \\ 
  &Tukey: ord & 2.0 & 0.07 \\ 
  &Tukey: log & 2.0 & 0.07 \\ 
  Qualitatively& Williams: 0 vs. 4 & 2.2 & 0.044 \\ 
 & Williams: 0 vs. 2+4 & 3.4 & 0.0013 \\ 
 & Williams: 0 vs. 1+2+4 & 4.6 & $<0.001$ \\ 
   \hline
\end{tabular}
\endgroup
\end{table}

\subsection{GLM and GLMM: overdispersed proportions}
Although most trend tests on proportions based on simple 2-by-k table data, studies with replicated proportions for each experimental unit within a group exist. Here overdispersion, i.e. variability between the individual proportions within a treatment group should be considered.
Several approach are available \cite{Hothorn2016}, here, quasi-binomial model in the GLM (using the quasibinomial(link="logit") link function)and a generalized linear mixed model with a random factor for the between-units variability are used.
As an example the germination rates for seed Orobanche cernua cultivated in three dilutions of a bean root extract in six replicates plates are used\cite{CROWDER1978}.

\begin{table}[ht]
\scriptsize
\centering
\caption{Example data with replicated proportions} 
\begin{tabular}{rrr}
  \hline
Dose & Un-germinated & Germinated \\ 
  \hline
1.000 &   43 &    2 \\ 
			&   51 &    9 \\ 
			&   44 &    5 \\ 
			&   71 &   16 \\ 
			&   24 &    2 \\ 
			&    7 &    0 \\ 
0.040 &   19 &   17 \\ 
			&   56 &   43 \\ 
			&   87 &   79 \\ 
			&   55 &   50 \\ 
			&   10 &    9 \\ 
0.002 &   13 &   11 \\ 
			&   62 &   47 \\ 
			&  104 &   90 \\ 
			&   51 &   46 \\ 
			&   11 &    9 \\ 
   \hline
\end{tabular}
\end{table}

The quasibinomial logit link function in the standard GLM can be used, where its
model is easily plugged into the \verb|tukeytrendfit| function. 

\footnotesize
\begin{verbatim}
oro<-glm(cbind(m,n) ~ dose, data=orob1, family=quasibinomial(link="logit"))
Exa17 <- tukeytrendfit(oro, dose="dose", scaling=c("ari", "ord", "log"))
EXA17<-summary(glht(model=Exa17$mmm, linfct=Exa17$mlf))
\end{verbatim}
\normalsize

Alternatively a generalized mixed effect model can be used, whereas the penalized quasi-likelihood approach; available via the function \verb|glmmPQL| is  particularly stable. The related code is more complex:

\footnotesize
\begin{verbatim}
OrobD <- dosescalett(orob1, dose="dose", scaling=c("ari", "ord", "log"))$data
glari <- glmmPQL(fixed=cbind(m,n) ~ doseari, random = ~ 1 |plate,
                   family = binomial, data=OrobD)
glord <- glmmPQL(fixed=cbind(m,n) ~ doseord, random = ~ 1 |plate,
                   family = binomial, data=OrobD)
gllog <- glmmPQL(fixed=cbind(m,n) ~ doselog, random = ~ 1 |plate,
                   family = binomial, data=OrobD)
lari <- tukeytrend:::lmer2lm(glari)
lord <- tukeytrend:::lmer2lm(glord)
llog <- tukeytrend:::lmer2lm(gllog)
linf <- matrix(c(0,1), ncol=2)
ttf <- summary(glht(mmm("mari"=lari, "mord"=lord, "mlog"=llog),
       mlf("mari"=linf, "mord"=linf, "mlog"=linf), alternative="less"))
							
\end{verbatim}
\normalsize

In either model, a strong significant decreasing germination rate according to the arithmetic dose score results.

\subsection{Further modifications}
Using both mmm and GLMM approaches, trend tests on i) multiple binary endpoints, ii) multiple differently scaled endpoints, iii) multinomial endpoints decomposed into correlated binary endpoints, iv) adjustments against covariate(s), v) non-monotonic dose-response relationships with a downturn effect at higher doses, vi) multi-center studies, vii) binary repeated measures, viii) cross-over designs are available, see related examples in the vignette \cite{vignette}.

\section{Conclusions}
By using the multiple marginal models approach (mmm), the widely used CA-trend test can be modified  which is powerful versus multiple shapes of dose response relationships (not just linear) for numerous special cases in GLMM for the three common effect sizes odds ratio, risk ratio and risk difference. \\
Examples for 2-by-k table data with a plateau shape and overdispersed proportions are discussed explicitly. More scenarios are given in a vignette to the R package \verb|tukeytrend| \cite{Schaarschmidt2018}.\\
The proposed approaches apply only asymptotically. In general, small sample sizes should generally be avoided for discrete endpoints, if only because of insufficient power.
\newline
\\
\textit{Acknowledgment: We thank Prof. Dr. Christian Ritz (University of Copenhagen) for the specific hints to the mmm approach in GLM and GLMM.}

\footnotesize
\bibliographystyle{plain}

\normalsize

\end{document}